\documentclass[twocolumn,noshowpacs,preprintnumbers,amsmath,amssymb,superscriptaddress]{revtex4}
\usepackage{graphicx}
\usepackage{graphicx,epsf} 
\usepackage{dcolumn}
\usepackage{bm}
\usepackage{latexsym}

\newcommand{\xsize}{\epsfxsize=7.0cm}
\newcommand{\xsec}{Appendix }

\usepackage{verbatim}

\begin{document}

\title{The Random Quadratic Assignment Problem}
\author{Gerald~Paul}
\affiliation{Center for Polymer Studies and Dept.\ of Physics, Boston
  University, Boston, MA 02215, USA} 
\email{gerryp@bu.edu}

\author{Jia~Shao}
\affiliation{Center for Polymer Studies and Dept.\ of Physics, Boston
  University, Boston, MA 02215, USA}

\author{H.~Eugene Stanley}
\affiliation{Center for Polymer Studies and Dept.\ of Physics, Boston
  University, Boston, MA 02215, USA} 
 

\begin{abstract}

  Optimal assignment of classes to classrooms \cite{dickey}, design of
  DNA microarrays \cite{carvalho}, cross species gene analysis
  \cite{kolar}, creation of hospital layouts \cite{elshafei}, and
  assignment of components to locations on circuit boards
  \cite{steinberg} are a few of the many problems which have been
  formulated as a quadratic assignment problem (QAP).  Originally
  formulated in 1957, the QAP is one of the most difficult of all
  combinatorial optimization problems.  Here, we use statistical
  mechanical methods to study the asymptotic behavior of problems in
  which the entries of at least one of the two matrices that specify
  the problem are chosen from a random distribution $P$.
  Surprisingly, this case has not been studied before using
  statistical methods despite the fact that the QAP was first proposed
  over 50 years ago \cite{Koopmans}.  We find  simple forms
  for $C_{\rm min}$ and $C_{\rm max}$, the costs of the minimal and
  maximum solutions respectively.    Notable features of our results
  are the symmetry of the results for $C_{\rm min}$ and $C_{\rm max}$
  and the dependence on $P$ only through its mean and standard
  deviation, independent of the details of $P$.  After the asymptotic
  cost is determined for a given QAP problem, one can straightforwardly
  calculate the asymptotic cost of a QAP problem specified with a
  different random distribution $P$.
\end{abstract}  

\maketitle


The quadratic assignment problem (QAP) is a combinatorial optimization
problem first introduced by Koopmans and Beckmann \cite{Koopmans}.  It is
NP-hard and is considered to be one of the most difficult problems to
be be solved optimally.  The problem was defined in the following
context: A set of $N$ facilities are to be located at $N$ locations.
The quantity of materials which flow between facilities $i$ and $j$ is
$A_{ij}$ and the distance between locations $i$ and $j$ is $B_{ij}$.
The problem is to assign to each location a single facility so as to
minimize (or maximize) the cost
\begin{equation}
C=\sum_{i=1}^N \sum_{j=1}^N  A_{ij} B_{p(i)p(j)},
\end{equation}
where $p(i)$ represents the location to which $i$ is assigned.

In addition to being important in its own right, the QAP includes such
other combinatorial optimization problems as the traveling salesman
problem and graph partitioning as special cases. There is an extensive
literature which addresses the QAP and is reviewed in
\cite{Pardalos,Cela,Anstreicher,Loiola,James,hanan,liggett,
  nagarajan,makarychev,burkard98,burkard09,commander}. With the
exception of specially constructed cases, optimal algorithms have
solved only relatively small instances with $N \le 36$. Various
heuristic approaches have been developed and applied to problems
typically of size $N\approx 100$ or less.

Most work on the QAP has focused on solution techniques, bounds on
optimal solutions, heuristics, and properties of problems with specially
structured matrices. Previous work on asymptotic properties of random
QAP instances has been limited to the case in which the elements of
{\it both} matrices are drawn from random distributions. In this case it was
shown  by rigorous arguments \cite{burkardFincke,frenk, albrecher2005,albrecher2006,krokhmal} that  almost surely
 as $ N \rightarrow \infty$, ($C_{\rm max} - C_{\rm min})/C_{\rm max} \to 0$; the minimum and maximum solutions approach the
solution obtained by a random permutation of $p$.

Here we consider the properties of solutions to the QAP under the
 requirement that the elements of only one of the matrices need
be drawn from random distribution $P$.  Our approach makes use of the
replica approach of statistical mechanics.

Without loss of generality, we will choose $A$ as a matrix the
elements of which are chosen from the random distribution P$(A_{ij})$;
the elements of B are arbitrary. We find that in the asymptotic limit
in which the size of the problem $ N \rightarrow \infty$
\begin{eqnarray}
C_{\rm min}&=&\mu_A \mu_B N^2  - \sigma_A f(B) N^{3/2} 
\label{optimal} \\
C_{\rm max}&=&\mu_A \mu_B N^2  + \sigma_A f(B) N^{3/2}.
\label {optimalMax}
\end{eqnarray}
Here $C_{\rm min}$ and $C_{\rm max}$ are the costs of the minimum and
maximum solutions, respectively, and $\mu_A$ and $\sigma_A$ are the
mean and standard deviations of the distribution $P(A)$; $\mu_B$ is the
mean of the entries of B and $b$ is a function of $B$ and $N$.  Our goal is to argue for the {\it form} of equations (\ref{optimal}) and (\ref{optimalMax}). We do
 not attempt to determine the value of the functions $f(B)$.


It is useful to first consider the solution for which $p$=$p^*$ is a
random permutation. Because the elements of $A$ are
assigned randomly and since $p^*(i)$ and $p^*(j)$ are random, each
 $B_{ij}$ in the sum is multiplied by a random value of $A_{ij}$ the
 average of which is $\mu_A$.  Hence, the cost of a random permutation
 is
\begin{eqnarray}
 C_{\rm rand}&=& \mu_A \sum_{i,j=1}^N    B_{p^*(i)p^*(j)}
=  \mu_A \mu_B N^2.
\label{eqRandom}
\end{eqnarray}


We now use the replica method of statistical mechanics to derive the
form for $C_{\rm min}$ and then derive the relationship of $C_{\rm max}$ to
$C_{\rm min}$. Employing a Hamiltonian, $\cal H$, defined as the QAP cost
function our goal is to compute the partition function
\begin{equation}
Z=\sum_{\{p\}} \exp[\frac{H}{kT }]= \sum_{\{p\}} \exp[ \frac{1}{kT } \sum_{i,j=1}^N A_{ij} B_{p(i)p(j)}   ]
\end{equation}
and the free energy
\begin{equation}
  -\frac{F}{kT}= \lim_{N \to \infty} \ln{Z}
\label{F}
\end{equation}
where $k$ and $T$ are the Boltzmann constant and temperature respectively.  Then, 
\begin{equation}
C_{\rm min}=F(T=0).
\label{cmin}
\end{equation}
Since the Hamiltonian includes a random matrix, $A$, we want to
calculate the value of the free energy $F$ averaged over the disorder
specified by the probability distribution $P(A)$.  However, averaging
the log of the partition function is difficult.  The replica method of
statistical mechanics \cite{EA} was introduced to make calculation of
this average possible.  The replica method has been used not just on
models of physical systems (such as spin glasses \cite{EA,sgtb,SK})
but also on such combinatorial optimization problems as graph
partitioning and the traveling salesman problem \cite{sgtb,Fu
  Anderson,hartmann}.  The calculation of the average of the partition
function is simplified using a mathematical identity known as the {\it
  replica trick}, $\ln(x) =\lim_{n \to 0}(x^n-1)/n$.  Then equation
(\ref{F}) becomes
\begin{equation}
  -\frac{F}{kT}= \lim_{N \to \infty}   \lim_{n \to 0}{1\over{n}}(\overline {Z^n}-1),            
\label{betaF}
\end{equation}
where
\begin{eqnarray}
Z^n&=&(\sum_{\{p^{1}\}} \exp{[\frac{ H(\{p^1\})}{kT}]})     \dots    (\sum_{\{p^{n}\}} \exp{[\frac{ H(\{p^n\})}{kT}]})  
\end{eqnarray}
and $\overline{Z^n} \equiv \int P(A) Z^n dA$ denotes $Z^n$ averaged over
the disorder. Here each Hamiltonian represents a replica of the
original system and the sum over ${\{p^{\alpha}\}}$ now denotes the
sum over all permutations in all replicas.


In order to achieve physically sensible results with $f\equiv F/N $
intensive, we require the following dependence on the mean and
standard deviation of P(A) to scale as  (see \cite{SK,Fu Anderson,sgtb}):
\begin{eqnarray}
\mu_A &=&\tilde \mu_A /N     \nonumber \\
\sigma_A&=&\tilde \sigma_A  /\sqrt{N}
\end{eqnarray}
%
%
with $\tilde \mu_A$ and $\tilde \sigma_A$ independent of $N$.  In \xsec 
A 
 we then find that 
\begin{eqnarray}
 \overline{Z^n} &= &   \exp[ \frac{n \mu_A  \mu_B  N^2}{kT}]  \nonumber \\
&\times&   \sum_{\{p^{\alpha}\}}  \exp \left[  \frac{\sigma_A^2}{2(kT)^2}  \sum_{i,j}   (\sum_{\alpha}^n  B_{p^\alpha(i)p^\alpha(j)} )^2 \right]
\label{zn10} 
\end{eqnarray}
%

We can make the following observations based on equations (\ref{zn10})  and (\ref{betaF}):
\begin{itemize}
\item Consistent with equation  (\ref{optimal}), the dependence of $F$ on
  $A$ is only through $\mu_A$ and $\sigma_A$.

\item If $\sigma_A=0$ and/or $T \to \infty$, substituting
  equation~(\ref{zn10}) in equation~(\ref{betaF}) yields  $F= \mu_A \mu_B N^2 $ which is the cost of the random
  solution, equation (\ref{eqRandom}).  This is reasonable because (i) physically for
  high temperatures we expect randomness and (ii) if
  $\sigma_A=0$, all entries in $A$ are identical and all permutations
  yield the same costs.
\end{itemize}
We infer the form of $F(T=0)$ as follows:
\begin{itemize}
\item In equation(\ref{zn10}) $\sigma_A$ appears in the combination $\sigma_A /T$.
  Thus, from equation (\ref{betaF}) we see that in the $T \to 0$ limit, only 
 a term linear in $\sigma_A$ can survive in $F$.

  This linear dependence on $\sigma_A$ as well as on $\mu_A$ is
  consistent with the simple case in which all of the
  elements in A are scaled by a constant, $z$, in which case $\sigma_A
  \rightarrow z \sigma_A$ and $\mu_A \rightarrow z \mu_A$.  Clearly
  the optimal permutation  is unchanged but the cost is also scaled by $z$.  Thus, in this
  simple case, for any permutation (including the optimal one) the
  linear dependence on $\mu_A$ and $\sigma_A$ must hold.

\item Given that we obtained equation (\ref{zn10}) by expanding in
  $1/\sqrt{N}$, we expect the second term in the expressions for
  $C_{\rm min}$ to be proportional $N^{3/2}$ since the leading term is
  proportional to $N^2$.

\end{itemize}
 Given the considerations, above the only possible
  expression for the second term in $F$ is $ \sigma_A f(B) N^{3/2}$ where
  $f$ is a function of $B$ only.


  The form of $C_{\rm max}$ follows directly as follows. Let $C_{\rm min}(A,B)$
  and $C_{\rm max}(A,B)$ denote the optimal minimum and maximum costs
  respectively of the QAP problem with matrices $A$ and $B$ and let
  $-A$ denote a matrix with elements $-A_{ij}$.  Since
  $C_{\rm max}(A,B)=-C_{\rm min}(-A,B)$ and since $\mu_{-A}=-\mu_{A}$ and
  $\sigma_{-A}= \sigma_{A}$, the form for $C_{\rm max}$ in equation
  (\ref{optimalMax}) follows directly from the form for $C_{\rm min}$.


If the entries of $B$ are also drawn from a random distribution, it is
straightforward to show that
\begin{eqnarray}
C_{\rm min}=\mu_A \mu_B N^2  - c \sigma_A \sigma_B N^{3/2}  \nonumber \\
C_{\rm max}=\mu_A \mu_B N^2  + c \sigma_A \sigma_B N^{3/2}
\label {optimal1}
\end{eqnarray}
where $c$ is a constant independent of $A$ and $B$ and $\sigma_B$ is
the standard deviation of the entries in $B$.


Zdeborov\'a et al. \cite{Zdeborova} have conjectured that, in the
large $N$ limit, the minimal and maximal costs of partitioning random
regular graphs into two equal sized subgraphs are related by $C_{\rm
  max}-\frac{|E|}{2}=\frac{|E|}{2} - C_{\rm min}$ where $|E|$ is the
total number of edges in the random regular graph.   Given that the
graph partitioning problem can be represented as a QAP (see \xsec
B)
this relationship follows directly from    equations (\ref{optimal})
and (\ref{optimalMax})  and supports this conjecture for the
partitioning of random graphs into two subgraphs of {\it any} size.  The more general 
 relationship   $C_{max} -\frac{\mu_A \mu_B N^2}{2}=-\frac{\mu_A \mu_B N^2}{2}-C_{min}$   can be interpreted as an extension to weighted as well as unweighted graphs.  


We are not aware of a method to proceed further with the replica
calculation for the case in which the $B$ matrix is not further
specified, but if the matrix elements $B_{ij}$ can be represented as
$b_i b_j$ (where $b_1, b_2,\dots b_N$ can take on  arbitrary values) the
problem is tractable and can be taken further.  This calculation is carried out in \xsec 
C
and yields a result in the form of equation (\ref{optimal}).


To confirm our findings, we use the tabu search (TS) \cite{Taillard}
heuristic to obtain approximate numerical solutions for a number of QAP
instances. We employ matrices of the types  described in detail in \xsec 
E.
 We use the notation {\it  "A matrix type"-"B matrix type"} to specify  a QAP instance.

In Fig. \ref{GG}, we plot $C_{\rm min}$ and $C_{\rm max}$ versus $\sigma_A$ for an instance of type Gaussian-Grid.  As expected, the plots are linear  in $\sigma_A$ and the absolute values of $C_{\rm min}$ and $C_{\rm max}$ are  equal for a given $\sigma$.  This is consistent with equations  (\ref{optimal}) and (\ref{optimalMax}).

A stronger test is achieved by studying  instances specified by a  matrix that represents a random graph of average degree $k$.  In this case,
\begin{equation}
\sigma_A(k)= \sqrt{ k(N-1-k)}=\sqrt{(\frac{N-1}{2})^2 - (k- \frac{N-1}{2})^2}
\label{circle}
\end{equation}
which represents a circle with origin at $( (N-1)/2,0)$.  In
Fig. \ref{ERG}(a) we plot $C_{\rm min}$, $C_{\rm rand}$, and $C_{\rm max}$
versus $k$, $0\leq k \leq N-1$, for an instance of type Random-Grid. In
order to illustrate the behavior of $C_{\rm min}$ and $C_{\rm max}$ in
more detail, in Fig.  \ref{ERG}(b), we plot
\begin{equation}
\Delta C_{\rm min/max} \equiv C_{\rm min/max} - C_{\rm rand}
\label{deltaC}
\end{equation}
The solid line is an ellipse of the form
\begin{equation}
C_{\rm max/min}^{\rm theory} =\pm\sigma_A(k) f(B) N^{3/2}
\label{th}
\end{equation}
where $f(B)$ is chosen to best fit of equation~(\ref{th}) to the data.  The
fit is consistent with the theory, exhibiting both the expected linear
dependence of the optimal costs on $\sigma_A(k)$ and the symmetry
represented by equations~(\ref{optimal}) and (\ref{optimalMax}).  In
Fig.
~\ref{var}.
 we show similar plots for other varied QAP instances.
Of particular interest are the plots for instances representing graph
partitioning; the plots illustrate the confirmation of the conjecture of
Ref.~\cite{Zdeborova} for both random and random regular graphs and
also the validity for partitioning of graphs into unequal sized sets of
vertices.

We now study the dependence of $\Delta C$ on $N$.  We treat instances
in which the $A$ matrix is random or random regular and consider
different types of $B$ matrix. To compare results for instances of
different sizes, we define the normalized quantities $\Delta C_{\rm norm}$
and $k_{\rm norm}$
\begin{eqnarray}
\Delta C_{\rm norm} \equiv \frac{\Delta C}{\mu_B N^2} \nonumber \\
k_{\rm norm} \equiv \frac{k}{N-1}.
\end{eqnarray}
With this normalization we expect 
\begin{equation}
\Delta C_{\rm norm} \sim \sigma_A N^{-1/2}
\label{cnorm1}.
\end{equation}
In  Fig. \ref{coll}(a) we plot $\Delta C_{\rm norm}$
for various values of $N$ for the Random-Grid instance.
We confirm the $N^{-1/2}$  dependence by plotting
\begin{equation}
\Delta C_{\rm collapsed} \equiv \Delta C_{\rm norm}N^{1/2}
\label{deltaCcoll}
\end{equation}
in  Fig. \ref{coll}(b).  The collapse is consistent with equation (\ref{cnorm1}). Additional plots for other instance types are shown in 
Fig.~ \ref{comb}.


In summary, using the replica method of statistical analysis, we have
found simple forms for the minimum and maximum costs of QAP problems
in which at least one matrix is determined by a random distribution.

We thank S. V.  Buldyrev for helpful discussions and the Defense
Threat Reduction Agency (DTRA) for support.


\begin{widetext}

\begin{figure}[!hp]

\centerline{
\xsize
\epsfclipon
\epsfbox{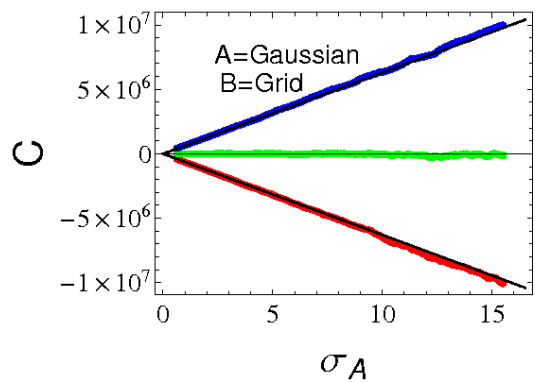}
}
\caption{For an $N=100$ QAP instance consisting of an $A$ matrix with
  elements from a Gaussian distribution and a $B$ matrix representing a
  two-dimensional grid, (from top to bottom), $C_{\rm max}$, $C_{\rm
    rand}$, and $C_{\rm min}$ versus standard deviation $\sigma_A$. For
  a given $\sigma_A$, $C_{\rm max}$ and $C_{\rm min}$ values are
  equidistant from $C_{\rm rand}$ value.  }
\label{GG}
\end{figure}

\begin{figure}[!hp]
\centerline{
\xsize
\epsfclipon
\epsfbox{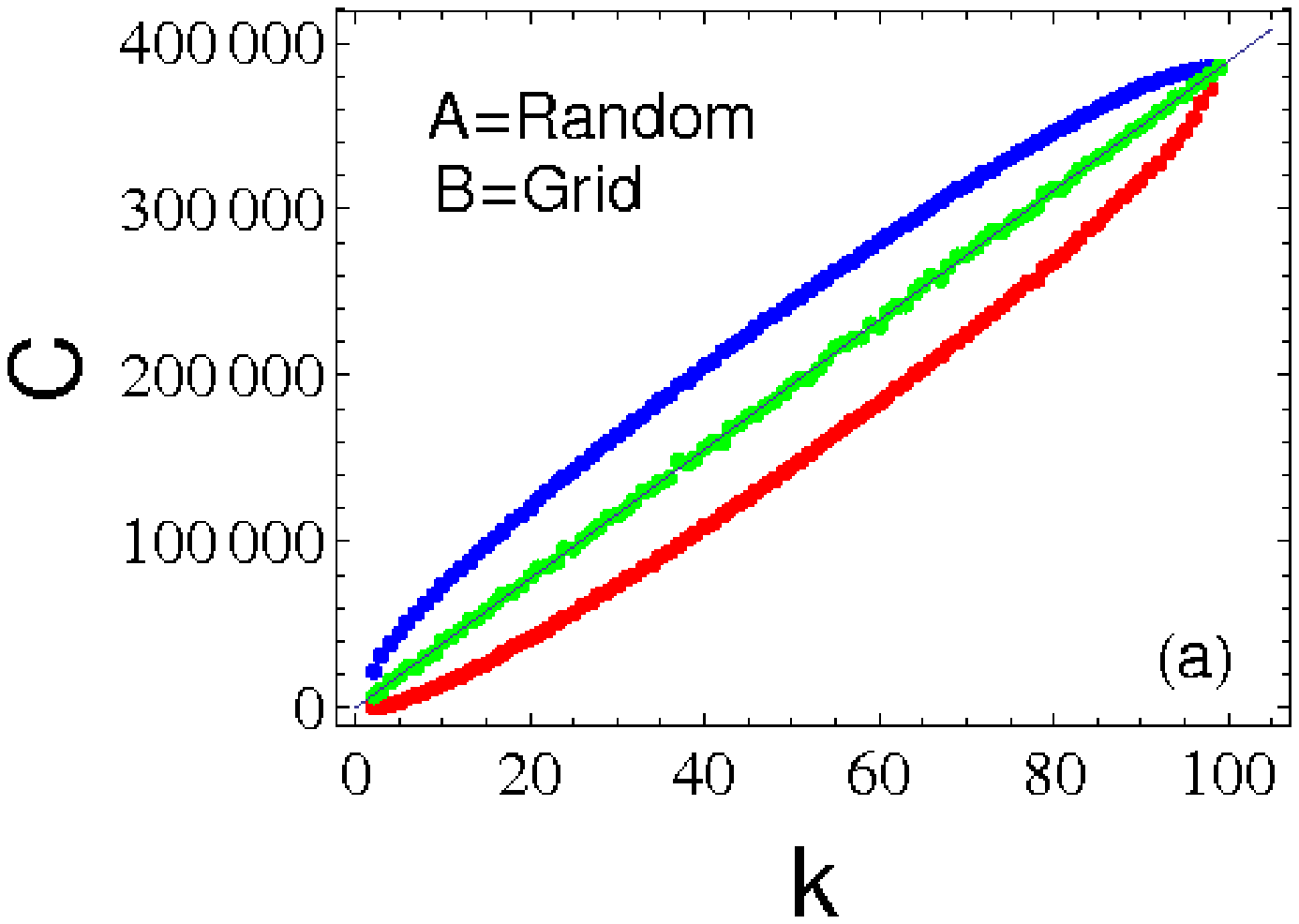}
}
\centerline{
\xsize
\epsfclipon
\epsfbox{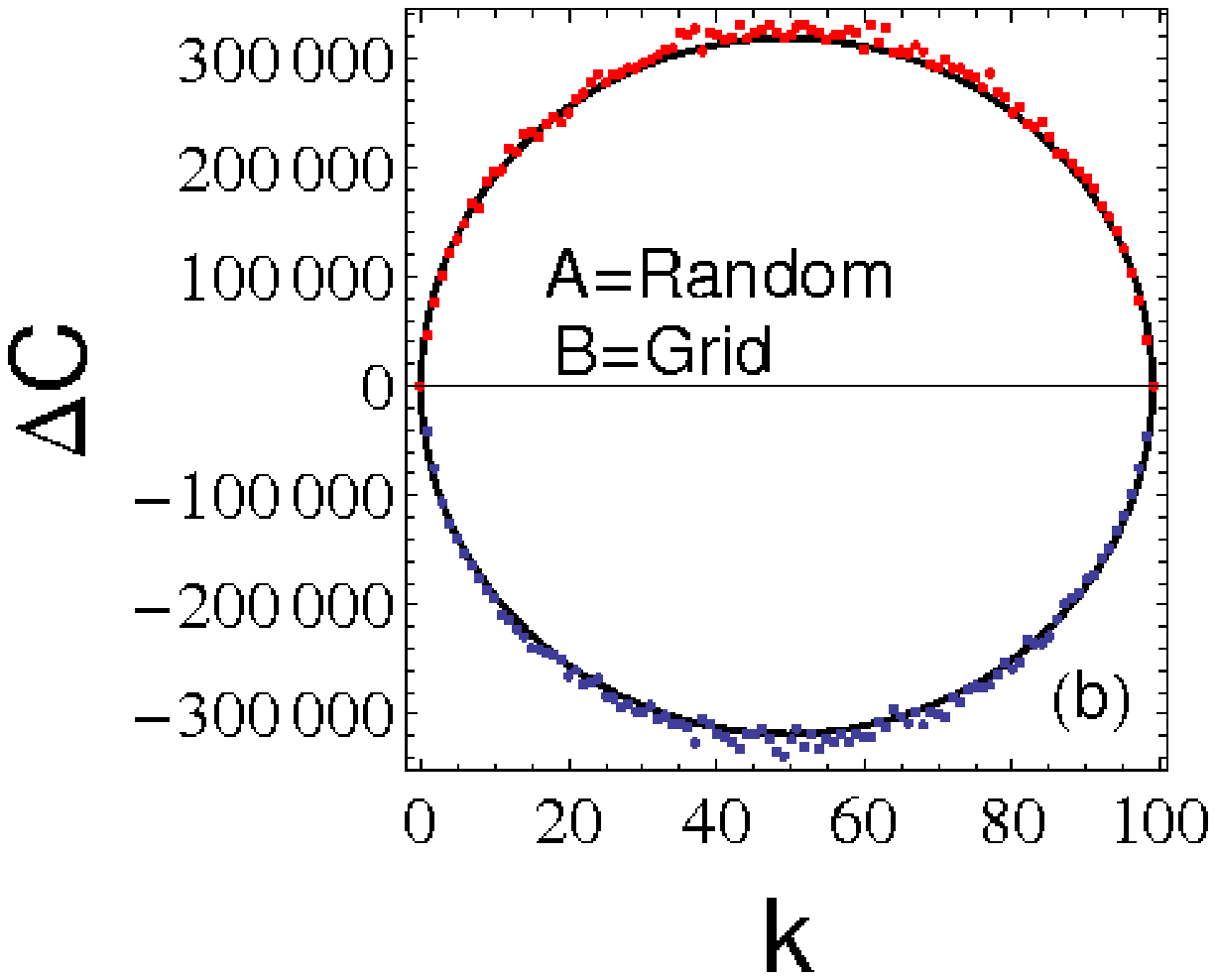}
}

\caption{ For an $N=100$ QAP instance consisting of an $A$ matrix
  representing a random graph and $B$ matrix representing a
  two-dimensional grid, (from top to bottom), (a) $C_{\rm max}$,
  $C_{\rm rand}$, and $C_{\rm min}$ versus average degree $k$ and (b) $\Delta
  C_{\rm max}$ and $\Delta C_{\rm min} $versus $k$.  In this and all following
  figures, the upper and lower semi-circles are the  $\Delta C_{\rm max}$
  and $\Delta C_{\rm min}$ plots, respectively.  The solid circular line
  represents the theoretical prediction.}
\label{ERG}
\end{figure}


\begin{figure}[h]
\begin{center}$
\begin{array}{cc}
\epsfxsize=7.0cm
\epsfbox{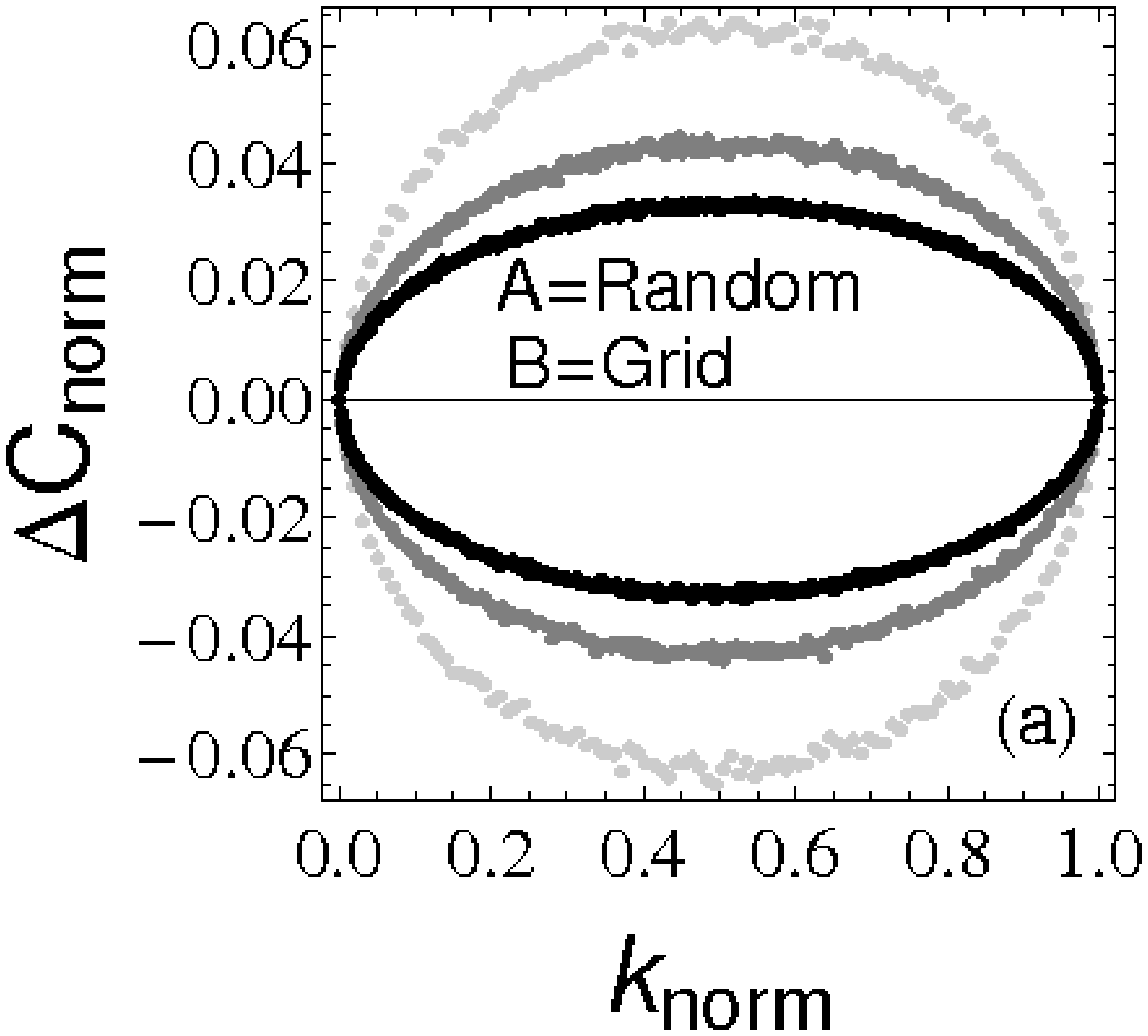} &
\epsfxsize=7.0cm
\epsfbox{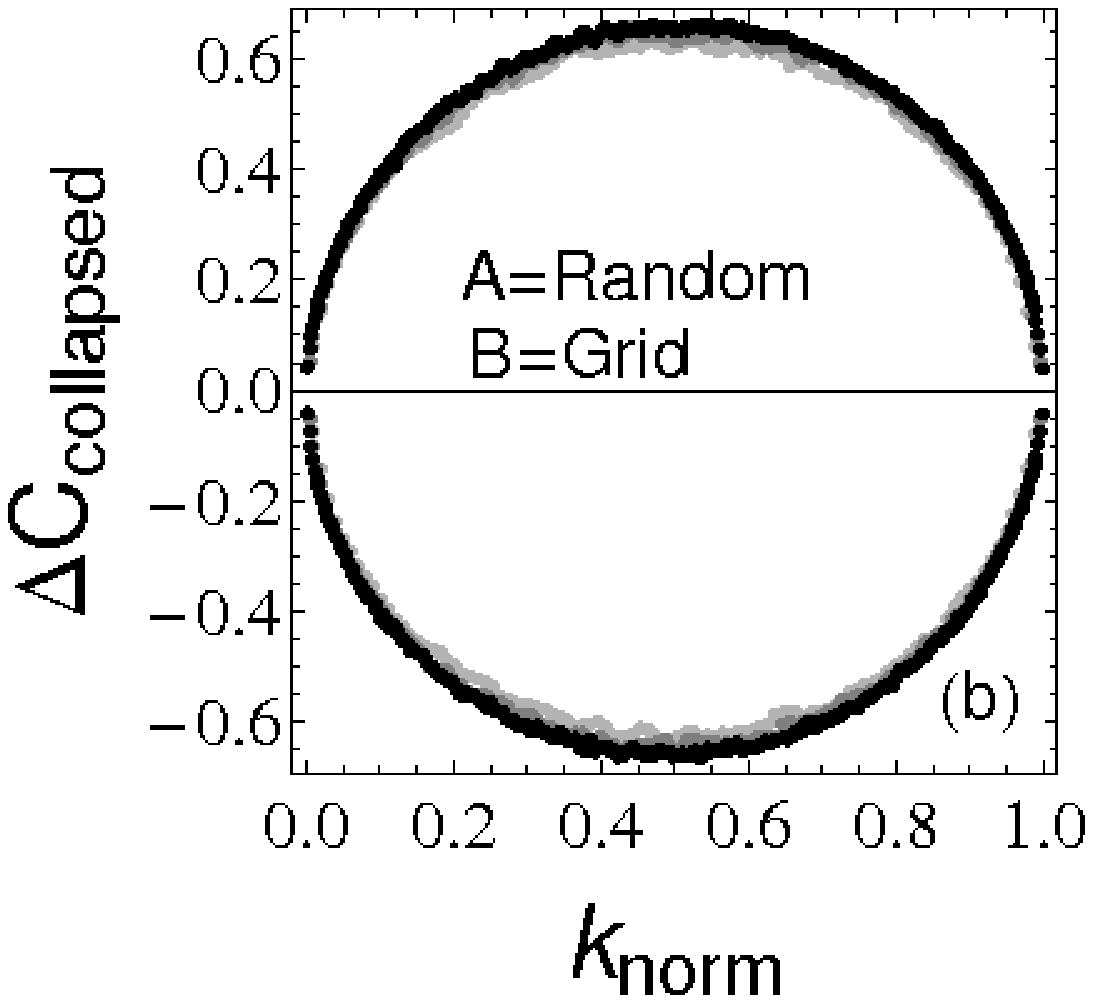}
\end{array}$
\end{center}
\caption{For a QAP instance of type Random-Grid, (a) normalized
  $\Delta C$ versus normalized $k$ for instance sizes $N$=100 (light
  gray); $ $N= 225 (medium gray), and $N$=400 (black); (b)
  corresponding collapsed plots (see equation (\ref{deltaCcoll}). }
\label{coll}
\end{figure}

\end{widetext}


\clearpage

\appendix

\section{Integration over disorder}
\label{disorder}

  In the following we retain only terms which do not vanish in the $N \to
\infty$ limit.  This is equivalent to retaining terms only to second
order in $A_{ij}$.  Because we want to maintain the exponential form,
we write

\begin{eqnarray}
\overline{Z^n} & =&\int P(A)   \sum_{\{p^{\alpha}\}} \exp{[\frac{1}{kT} \sum_{\alpha=1}^n    \sum_{i,j=1}^N A_{ij} B_{p^\alpha(i)p^\alpha(j)}   ]} dA      \nonumber \\     
&=& \sum_{\{p^{\alpha}\}}   \exp{[\ln\int P(A) e^{ \frac{1}{kT} \sum_{i,j}  \sum_{\alpha}^n   A_{ij} B_{p^\alpha(i)p^\alpha(j)}}  dA]} \nonumber \\
&=&  \sum_{\{p^{\alpha}\}}  \exp{[\ln \prod_{i,j}  \int P(A_{ij})  e^{  \frac{1}{kT} \sum_{\alpha}^n   A_{ij} B_{p^\alpha(i)p^\alpha(j)}}  dA_{ij}] } \nonumber \\
&=&   \sum_{\{p^{\alpha}\}}  \exp{[\sum_{i,j} \ln \int P(A_{ij})  e^{ \frac{1}{kT}  \sum_{\alpha}^n   A_{ij} B_{p^\alpha(i)p^\alpha(j)}}   dA_{ij}] } \nonumber \\
%
%
&=&   \sum_{\{p^{\alpha}\}}  \exp{[\sum_{i,j} \ln(1+y_{ij})] }  
\end{eqnarray}
where $y_{ij} \equiv   \int P(A_{ij})   \exp[{ \frac{1}{kT} \sum_{\alpha}^n A_{ij} B_{p^\alpha(i)p^\alpha(j)}}] dA_{ij} - 1 $ 
Expanding $y_{ij}$ to second order in $A_{ij}$ we have:
\begin{widetext}
\begin{eqnarray}
y_{ij} & \sim& \int P(A_{ij})[1+ \frac{1}{kT}  \sum_{  \alpha}^n    A_{ij} B_{p^\alpha(i)p^\alpha(j)} + \frac{(\sum_{\alpha}^n    A_{ij} B_{p^\alpha(i)p^\alpha(j)})^2}{2(kT)^2  }]dA_{ij} - 1  \nonumber \\
&=&   \frac{\mu_A}{kT} \sum_{\alpha}^n   B_{p^\alpha(i)p^\alpha(j)} + \frac{ \mu_{2A}}{2(kT)^2} (\sum_{\alpha}^n  B_{p^\alpha(i)p^\alpha(j)} )^2  \nonumber \\
\label{y1}
\end{eqnarray}
where  $\mu_{A2}$ is the second moment(around zero) of $P$. Expanding $\ln(1+y_{ij})$ to the second
order in $y_{ij}$ and substituting equation~(\ref{y1}), we have

\begin{eqnarray}
  \ln(1+y_{ij}) & \sim & \frac{ \mu_A}{kT} \sum_{\alpha}^n   B_{p^\alpha(i)p^\alpha(j)} + \frac{ \mu_{2A}}{2(kT)^2} (\sum_{\alpha}^n  B_{p^\alpha(i)p^\alpha(j)})^2  -\frac{1}{2}  (\frac{ \mu_A}{kT} \sum_{\alpha}^n   B_{p^\alpha(i)p^\alpha(j)} )^2    \nonumber \\
%
%
  &=&  \frac{ \mu_A}{kT} \sum_{\alpha}^n   B_{p^\alpha(i)p^\alpha(j)} + \frac{\sigma_A^2}{2(kT)^2} (\sum_{\alpha}^n  B_{p^\alpha(i)p^\alpha(j)} )^2 \nonumber \\     
\end{eqnarray}
where we have only retained terms to $O(A_{ij}^2)$.  Finally we have
\begin{eqnarray}
\overline{Z^n}&=&    \sum_{\{p^{\alpha}\}}   \exp{\left[\sum_{i,j}[  \frac{ \mu_A}{kT} \sum_{\alpha}^n   B_{p^\alpha(i)p^\alpha(j)} +\frac{\sigma_A^2}{2(kT)^2} (\sum_{\alpha}^n  B_{p^\alpha(i)p^\alpha(j)} )^2]\right] }  \nonumber  \\
&= &    \sum_{\{p^{\alpha}\}}    \exp\left[ \frac{\mu_A  \mu_B n N^2}{kT}   + \frac{\sigma_A^2}{2(kT)^2}  \sum_{i,j}   (\sum_{\alpha}^n  B_{p^\alpha(i)p^\alpha(j)} )^2 \right] 
\end{eqnarray}
\end{widetext}
where we  use the fact that   $\sum_{i,j}  B_{p^\alpha(i)p^\alpha(j)}   $  is independent of permutation.

\section { Relationship to Graph Partitioning}
\label{gp}

The problem of partitioning a graph into two subgraphs of size
$rN$ and $(1-r)N$ with the minimum number of edges between the two
subgraphs can be represented as a QAP as follows: One matrix, $A$, is
the adjacency matrix of the graph to be partitioned.  The other
matrix, $B$, the {\it graph partitioning matrix}, is the adjacency matrix for a bipartite graph in which
edges are present between two sets of vertices; one set contains $rN$
vertices and the second set contains $(1-r)N$ vertices.  The QAP
cost function is the cost of partitioning the graph represented by $A$.

\section{Specific B Matrix}
\label{specB}

Here we treat the case in which the matrix elements $B_{ij}$ can be
represented as $b_i b_j$.  We follow  the spin-glass  calculation of  Ref. \cite{SK}.

 Let
\begin{equation}
Z_n ' \equiv    \sum_{\{p^{\alpha}\}}     \exp[ \frac{ \tilde \sigma_A^2 N}{2 (kT)^2}  \sum_{i,j=1}^N (\sum_{\alpha=1}^n  B_{p^\alpha(i)p^\alpha(j)}/N)^2]
\label{zn11} 
\end{equation}
so
\begin{equation}
\overline{Z^n}=\exp[\frac{\mu_A \mu_B n N^2}{kT}] Z_n'.
\label{zn12}
\end{equation}
Using
\begin{eqnarray}
& &   \sum_{i,j=1}^N (\sum_{\alpha=1}^n b_{p^\alpha(i)} b_{p^\alpha(j)})^2 \nonumber \\
&=&\sum_{\alpha\beta=1}^n   (\sum_{i=1}^N    b_{p^\alpha(i)}     b_{p^\beta(i)} )^2  \nonumber \\
&=&  n N^2 \mu_{2b}^2 +  \sum_{(\alpha \neq \beta)=1}^n   (\sum_{i=1}^N    b_{p^\alpha(i)}     b_{p^\beta(i)} )^2, 
\end{eqnarray}
where $\mu_{2b}$ is the second moment of the elements of the vector $b$,  equation~(\ref{zn11}) becomes
\begin{eqnarray}
Z_n' &=& \sum_{\{p^{\alpha}\}}  \exp[ \frac{ \tilde \sigma_A^2 N}{2 (kT)^2}  (n  \mu_{2b}^2   + \sum_{(\alpha \ne\beta)=1}^n   (\sum_{i=1}^N \frac{  b_{p^\alpha(i)} b_{p^\beta(i)}}{N} )^2)  ]    \nonumber
\label{zn20}
\end{eqnarray}

We can now use the Gaussian integral identity
\begin{equation}
e^{\lambda z^2}= \frac{1}{\sqrt{2 \pi}} \int (-\frac{1}{2}x^2 + (2 \lambda)^{1/2}z x) dx   \nonumber
\end{equation}
with $dx \to  (\frac{\tilde \sigma_A^2 N}{2 (kT)^2})^{1/2}  dQ_{\alpha \beta}$ and find

\begin{widetext}

\begin{eqnarray}
Z_n'=   \exp\left[\frac{\tilde \sigma_A^2 N}{2 (kT)^2} (n \mu_{2b}^2 )\right]   \int \prod_{\alpha \beta}    ( \frac{ N \tilde \sigma_A^2 }{{2 \pi} (kT)^2 })^{1/2}  dQ_{\alpha \beta}
\exp \left[-N D[Q_{\alpha \beta}] \right] 
\label{zx1}
\end{eqnarray}
where 
\begin{equation}
D[Q_{\alpha \beta}] =\frac{ \tilde \sigma_A^2 }{2 (kT)^2}  \sum_{(\alpha \ne\beta)=1}^n Q_{\alpha \beta}^2  -    \ln{     \sum_{\{p^{\alpha}\}}     \exp  \frac{1}{2}  (    \frac{ \tilde \sigma_A}{ k T})^2 Q  \sum_{(\alpha \ne\beta)=1}^n   \sum_{i=1}^N    b_{p^\alpha(i)}      b_{p^\beta(i)} /N  }.
\end{equation}

The form of equation (\ref{zx1}) suggests that the integrals be evaluated
with the method of steepest descent with the value of the integral
determined by the maxim value of $D$.  Assuming no replica symmetry
breaking, at the maximum all values of $Q_{\alpha \beta}$ are equal
\cite{sgtb}; we denote this maximum value as $Q$ and
\begin{equation}
D[Q]= -(\frac{ \tilde \sigma_A}{2 k T})^2 n(n-1) Q^2 - \ln{     \sum_{\{p^{\alpha}\}}     \exp  \frac{1}{2}     (\frac{ \tilde \sigma_A}{ k T})^2  Q \sum_{(\alpha \ne\beta)=1}^n    \sum_{i=1}^N    b_{p^\alpha(i)}      b_{p^\beta(i)} /N  }
\end{equation}
Using
\begin{eqnarray}
&&\sum_{(\alpha \ne\beta)=1}^n   \sum_{i=1}^N    b_{p^\alpha(i)}      b_{p^\beta(i)}   \nonumber   \\
&=&      \sum_{i=1}^N       [ \sum_{\alpha}^n  ( b_{p^\alpha(i)})^2  -      \sum_{\alpha}^n    b_{p^\alpha(i)}^2]     \nonumber       \\
&=&   \sum_{i=1}^N        \sum_{\alpha}^n  ( b_{p^\alpha(i)})^2  -     n N \mu_{2b}  
\end{eqnarray}
to uncouple the replicas, we have 
\begin{equation}
D[Q]= -(\frac{ \tilde \sigma_A}{2 k T})^2 ( n(n-1) Q^2  +2 n  \mu_{2b} Q)         - \ln{     \sum_{\{p^{\alpha}\}}     \exp (\frac{ \tilde \sigma_A}{ k T})^2   \frac{1}{2}     \sum_{i=1}^N ( \sum_{\alpha}^n      b_{p^\alpha(i)} )^2  Q/N }
\end{equation}
In the spin glass calculation the sum over $i$ can be carried out here and the factor of $1/N$  drops out.  Since we cannot perform the sum over $i$ here, the $1/N$ factor is carried through to the end of the calculation as a normalization factor.   Using the Gaussian integral identity again we can write
\begin{eqnarray}
E[Q]&\equiv &   \ln{     \sum_{\{p^{\alpha}\}}     \exp[ (\frac{ \tilde \sigma_A}{ k T})^2   \frac{1}{2}     \sum_{i=1}^N ( \sum_{\alpha}^n      b_{p^\alpha(i)} )^2  Q/N }]  \nonumber  \\
&=&  \ln     \sum_{\{p^{\alpha}\}} \prod_i ^N \int  \exp[{-\frac{z_i^2}{2}  +  \frac{ \tilde \sigma_A}{ k T} \sqrt{Q/N} \sum_{\alpha}^n  b_{p^\alpha(i)}    z_i  }]      \frac{ dz_i}{\sqrt{2 \pi}} 
\end{eqnarray}
Now considering just the terms in the exponent  dependent on $b_{p^\alpha(i)}$, we find
\begin{equation}
\sum_{\{p^{\alpha}\}}  \exp[{ \frac{ \tilde \sigma_A}{ k T} \sqrt{Q/N} \sum_{\alpha}^n b_{p^\alpha(i)}    z_i  }]   
= \prod_\alpha ( \sum_{p^{\alpha}}  \exp[{ \frac{ \tilde \sigma_A}{ k T} \sqrt{Q/N} b_{p^\alpha(i)}    z_i  }])
= (\sum_{p}  \exp[{ \frac{ \tilde \sigma_A}{ k T} \sqrt{Q/N}\sum_{\alpha}^n  b_{p(i)}    z_i  } ]   )^n 
\end{equation}
This is the key step which makes the $n$ dependence explicit and allows non-ambiguous analytic continuation of $n \to 0$.   Keeping only terms to first order in $n$, and using $   (1/2 \pi) \int_{-\infty}^\infty e^{-z^2/2} dz=1$ we can then write  
\begin{eqnarray}
E[Q]&=&   \ln     \prod_i ^N \int   e^{-{z_i^2/2}}  ( \sum_p  \exp[{  \frac{ \tilde \sigma_A}{ k T} \sqrt{Q/N} b_{p(i)}    z_i] )^n }       \frac{ dz_i}{\sqrt{2 \pi}}) \nonumber \\
&=&    \ln     \prod_i ^N \int   e^{-{z_i^2/2} }   (1+ n \ln    \sum_p  \exp[{  \frac{ \tilde \sigma_A}{ k T} \sqrt{Q/N} b_{p(i)}    z_i   }]) \frac{ dz_i}{\sqrt{2 \pi}}) \nonumber   \\
&=&   \ln(1 +  n   \prod_i ^N \int   e^{-{z_i^2/2} }   \ln    \sum_p  \exp[{  \frac{ \tilde \sigma_A}{ k T} \sqrt{Q/N} b_{p(i)}    z_i  ] } \frac{ dz_i}{\sqrt{2 \pi}}) \nonumber  \\
&=&     n    \prod_i ^N \int   e^{-{z_i^2/2} }   \ln    \sum_p  \exp[{  \frac{ \tilde \sigma_A}{ k T} \sqrt{Q/N} b_{p(i)}    z_i ]  } \frac{ dz_i}{\sqrt{2 \pi}}
\end{eqnarray}
For small $T$,  as shown in \xsec \ref{EQappendix},
\begin{equation}
E[q]= n  \frac{ \tilde \sigma_A}{ k T}\sqrt{Q/N} \Psi(b),
\end{equation}
where $\Psi(b)$ depends only on the values of $b_i$ normalized by $\sqrt{N}$.  We can then write
\begin{equation}
D[Q]= -(\frac{ \tilde \sigma_A}{2 k T})^2  (  n(n-1) Q^2  +2 n \mu_{2b} Q ) -  n  \frac{ \tilde \sigma_A}{ k T}\sqrt{Q}\Psi(b)  
\label{DQ} 
\end{equation}

To find the Q which maximizes Z' we differentiate equation (\ref{DQ}) with respect to Q and solve for Q.  In the limit $n \to 0$ we find
\begin{equation}
Q= \mu_{2b} - \frac{k T}{\tilde \sigma_A} \Psi.
\label{Q}
\end{equation}

Substituting for $D[Q]$ and evaluating equation (\ref{zx1}) using the method of steepest descent we find, to first order in $n$,
\begin{eqnarray}
Z_n'&= &   \exp[   - N  (\frac{ \tilde \sigma_A}{2 k T})^2  (   - \mu_{2b}^2 n+  n(n-1) Q^2  +2 n \mu_{2b} Q ) -  n  \frac{ \tilde \sigma_A}{ k T}\sqrt{Q}\Psi(b)          ] \nonumber \\
&=& \exp[n N (   (\frac{ \tilde \sigma_A}{2 k T})^2  (    (\mu_{2b} -  Q )^2 -  \frac{ \tilde \sigma_A}{ k T}\sqrt{Q}\Psi(b) )          ]
\label{zx2}
\end{eqnarray}
and thus from equation(\ref{zn12}) 
\begin{eqnarray}
F(T=0)&=& -k T \lim_{n \to 0}{1\over{n}}(\overline{Z^n}-1) \nonumber \\ 
&=&  - k T[ \mu_A \mu_B N^2  +     (\frac{ \tilde \sigma_A}{2 k T})^2 N  (    (\mu_{2b} -  Q )^2 -  \frac{ \tilde \sigma_A
}{ k T}\sqrt{Q}\Psi(b) ) ]
\end{eqnarray}

Using equation (\ref{Q}), in the limit $T \to 0$, and defining  $f(b)=\sqrt{ \mu_{2b}} \Psi(b)$
\begin{eqnarray}
F(T=0)&=&\mu_A \mu_B N^2 -  \tilde \sigma_A N f(b) \nonumber \\
 &=&\mu_A \mu_B N^2 -  \sigma_A N^{3/2} f(b) 
\end{eqnarray}. 
%

\section{E[Q] calculation}
\label{EQappendix}

Here we discuss the evaluation of $E[Q]$.  We want to evaluate multiple integrals of the form
\begin{equation}
  E[Q]= \prod_i ^N \int   e^{-{z_i^2/2} }   \ln    \sum_p  \exp[{  \frac{ \tilde \sigma_A}{ k T} \sqrt{Q/N} b_{p(i)}    z_i ]  } \frac{ dz_i}{\sqrt{2 \pi}}.
\end{equation}
To see how the evaluation would proceed, consider the case of N=2.  Then
\begin{equation}
E[Q]= \int  e^{-{z_1^2/2} }  (    \int   e^{-{z_2^2/2} } \ln  [e^{a(b_1 z_1 + b_2 z_2)} +  e^{a(b_2 z_1 + b_1 z_2}]
   \frac{ dz_1}{\sqrt{2 \pi}}) \frac{ dz_2}{\sqrt{2 \pi}},
\end{equation}
where $a \sim 1/T $ .  For small $T$, for different regions of integration, one exponential in the sum of exponentials dominates.  Specifically, assuming without loss of generality that $b_1 > b_2$,
\begin{eqnarray}
 \ln  [e^{a(b_1 z_1 + b_2 z_2} +  e^{a(b_2 z_1 + b_1 z_2}] &\sim& a b_1 z_1 +\ln[ e^{a  b_2 z_2}] ~~~~~~ (z_1>z_2  )   \nonumber \\
                                                                                                   &\sim& a b_2 z_1 +\ln[ e^{a  b_1 z_2}] ~~~~~~ (z_1<z_2  ).
\end{eqnarray} 
Then
\begin{eqnarray}
  E[Q]= \int  e^{-{z_1^2/2} }  (    \int_{z_2}^\infty   e^{-{z_2^2/2}}    a b_1 z_1 +\ln[ e^{a  b_2 z_2}] dz1 + \int_{-\infty}^{z_2}      a b_2 z_1 +\ln[ e^{a  b_1 z_2}] ]   \frac{ dz_1}{\sqrt{2 \pi}}) \frac{ dz_2}{\sqrt{2 \pi}}.
\end{eqnarray}
These integrals can all be solved exactly, with intermediate results
in terms of the error function, $ erf(x)=\frac{2}{\sqrt{\pi}}\int_0^x
e^{-z^2} dz$, and we find that for this case of $N=2, E[Q]= a(b_1 -
b_2)/\sqrt{\pi}$.  This approach can be extended to any $N$ and the
result is $a$ times a linear combination of the constants $b_i$.  We
can then write
\begin{equation}
E[q]= n  \frac{ \tilde \sigma_A}{ k T}\sqrt{Q} \Psi(b),
\end{equation}
where $\Psi(b)$ is the linear combination of the $b_i$ divided by $\sqrt{N}$ which normalizes the expression.

\end{widetext}

\section{Matrix Types}
\label{mt}

We employ matrices of the following types:

\begin{itemize}
 
\item Uniform - the matrix elements are chosen from a uniform
  distribution on the interval $[0,100]$.

\item Gaussian - matrix elements are chosen from a Gaussian
  distribution with zero mean and standard deviation $\sigma$.

\item Half-Gaussian - matrix elements are chosen from a Gaussian
  distribution as above but only elements with value greater or equal
  to zero are used.

\item Random (graph) - the matrix is the adjacency matrix of a random graph
  with edges present with probability $p$.  The average degree of the
  graph is $k=p N$.   

\item Random Regular (graph) - the matrix is the adjacency matrix of a random
  regular graph for which all vertices are degree $k$.

\item Grid - the matrix elements are the Euclidean distances between
  points in a two-dimensional square grid.  The distances between
  adjacent points along the $x$ and $y$ axes are $100$.

\item Graph Partitioning - the matrix is the graph partitioning matrix described in \xsec \ref{gp}.

\end{itemize}
All matrices are symmetrical with zero diagonal.  For the Random and
Random Regular matrices that represent graphs, we study cases of the
graph degree $k$ ranging from $0$ to $N-1$. 
%

\begin{widetext}

\begin{figure}[h]
\begin{center}$
\begin{array}{cc}
\epsfxsize=7.0cm
\epsfbox{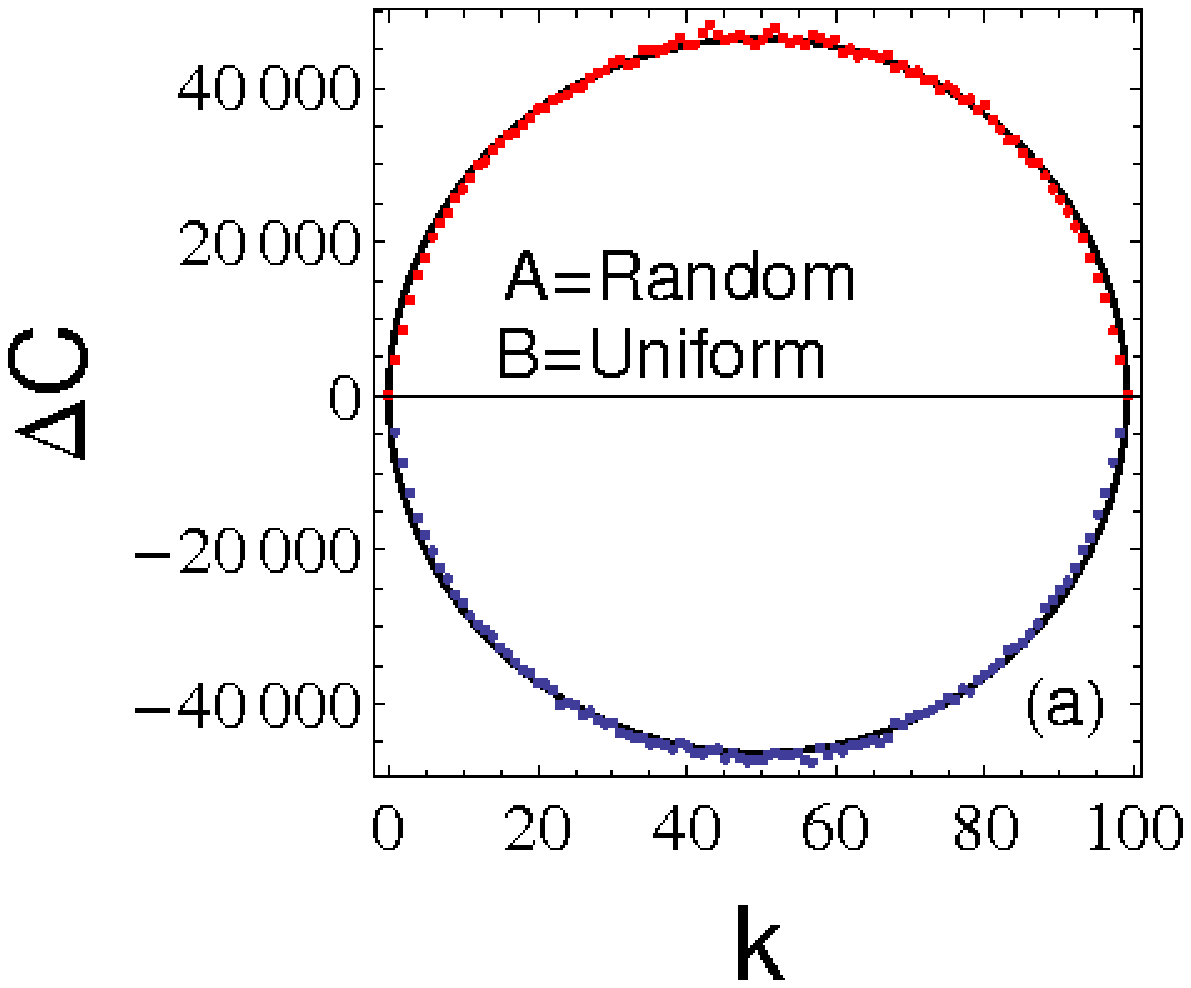} &
\epsfxsize=7.0cm
\epsfbox{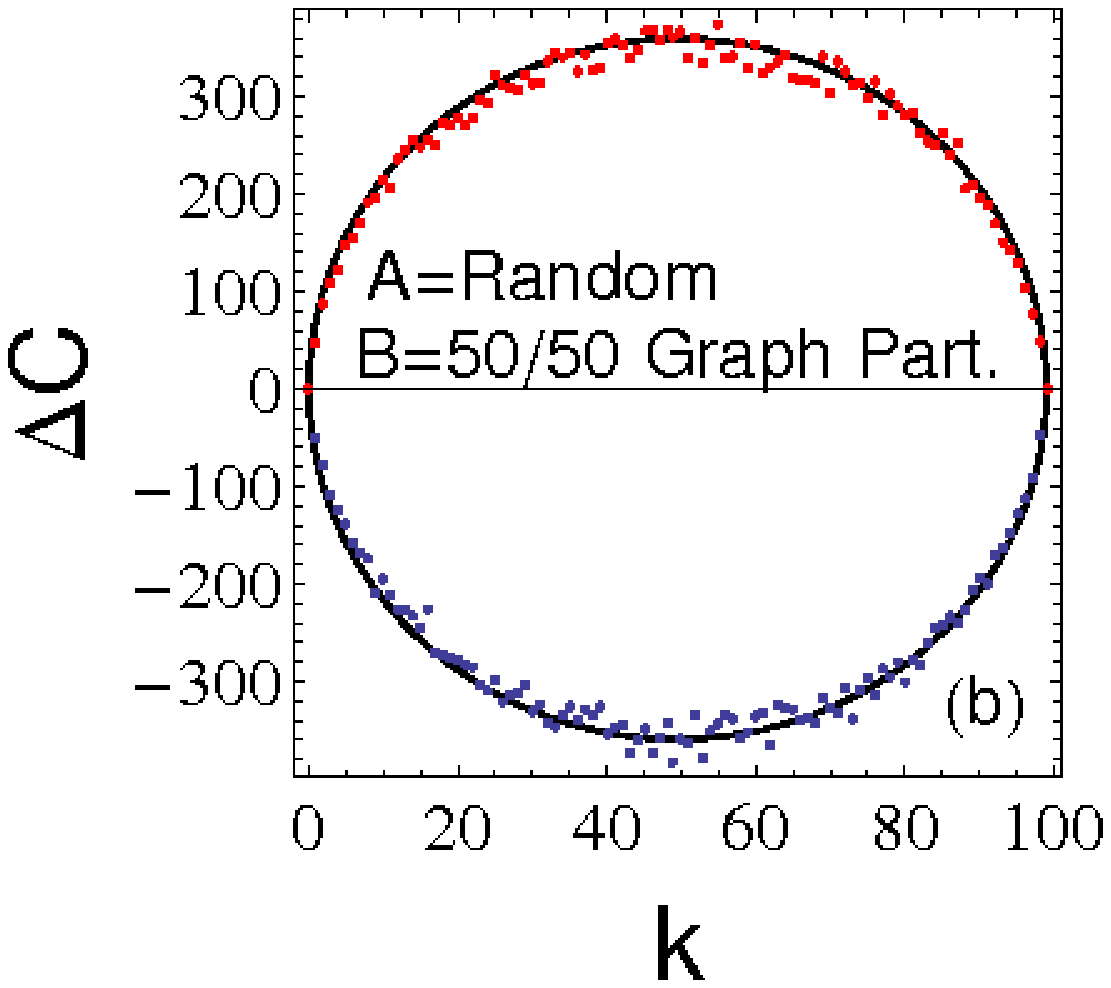}\\
\epsfxsize=7.0cm
\epsfbox{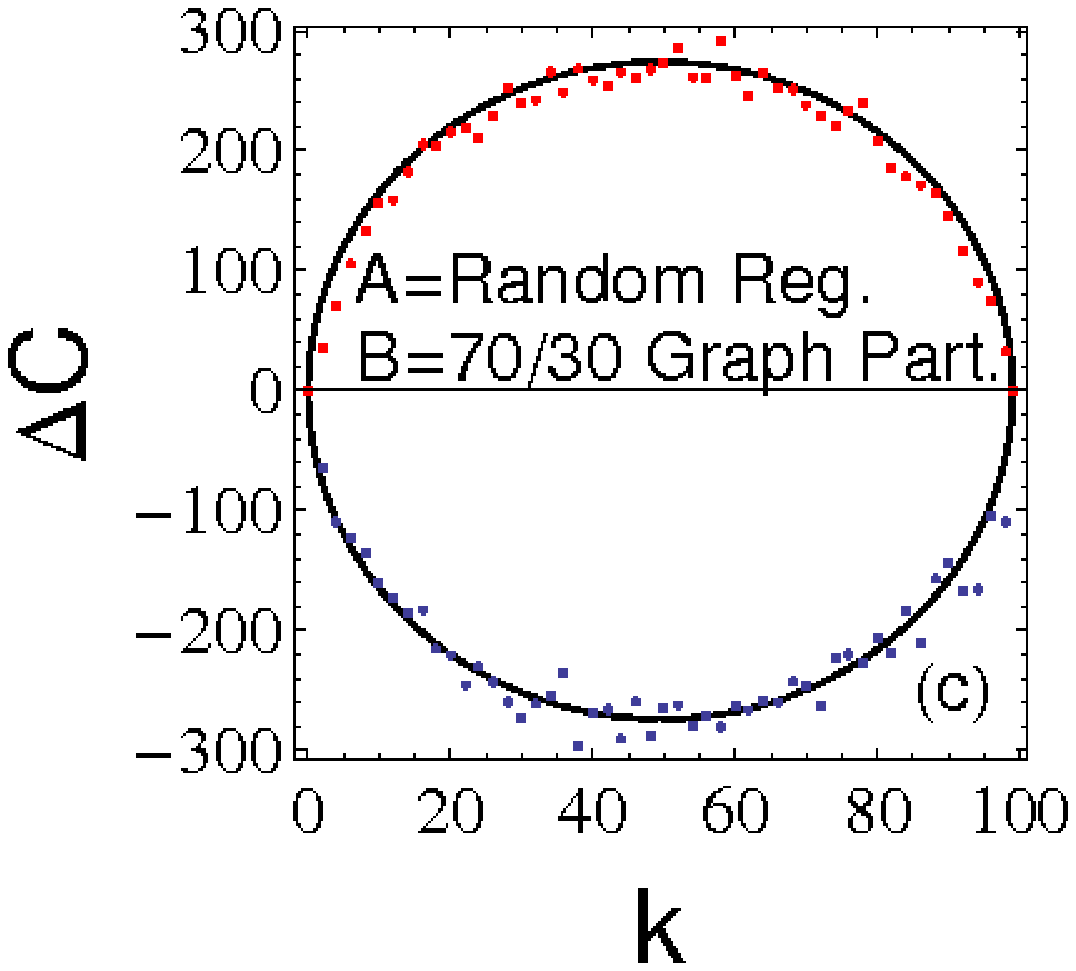} &
\epsfxsize=7.0cm
\epsfbox{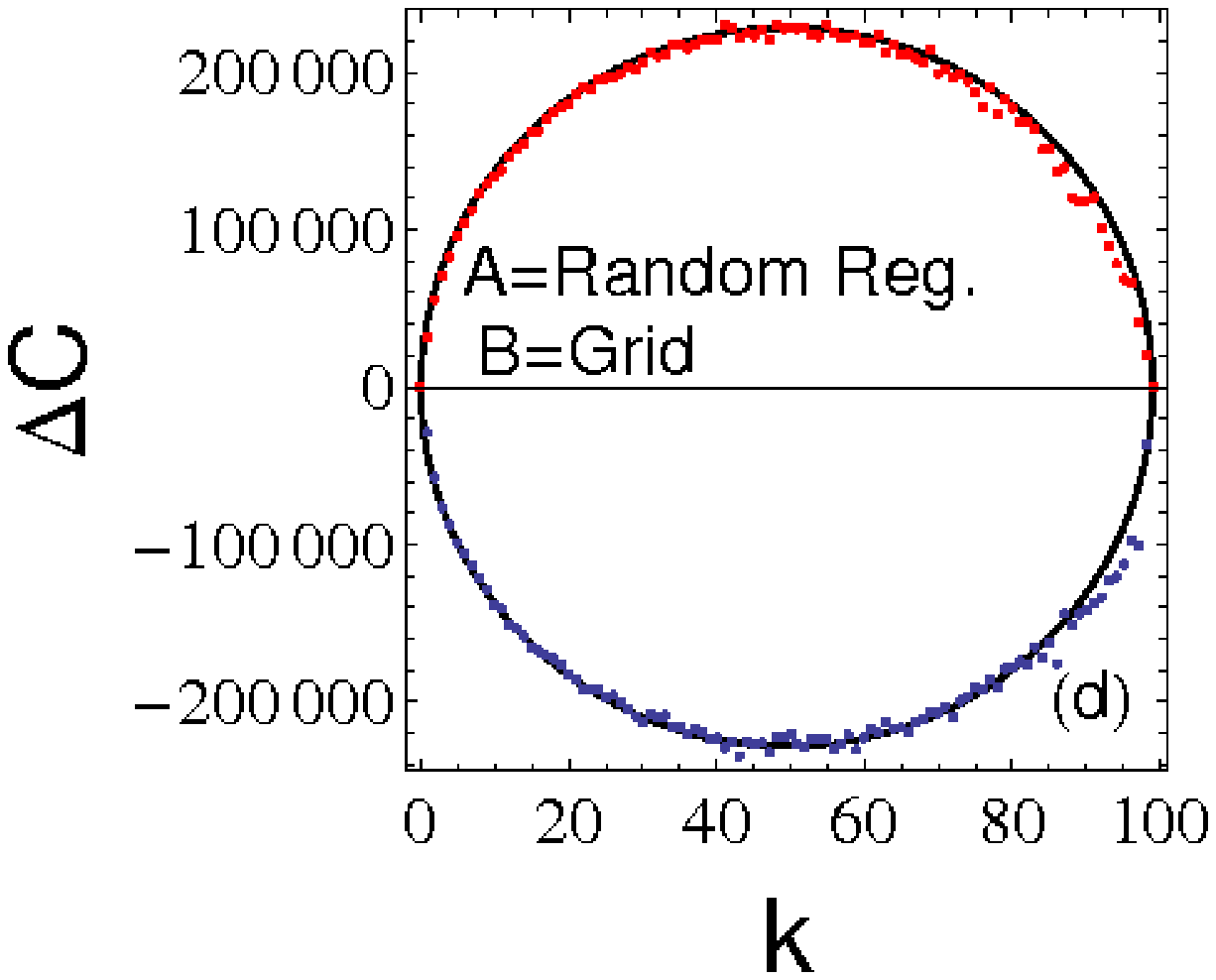}
\end{array}$
\end{center}
\caption{  For various  $N=100$ QAP instance,  $\Delta C_{\rm max}$ and $\Delta C_{\rm min} $versus $k$.  The solid circular line represents the theoretical prediction.    }
\label{var}
\end{figure}


\begin{figure}[h]
\begin{center}$
\begin{array}{cc}
\epsfxsize=7.0cm
\epsfbox{pCombERGa.eps} &
\epsfxsize=7.0cm
\epsfbox{pCombERGb.eps}\\
\epsfxsize=7.0cm
\epsfbox{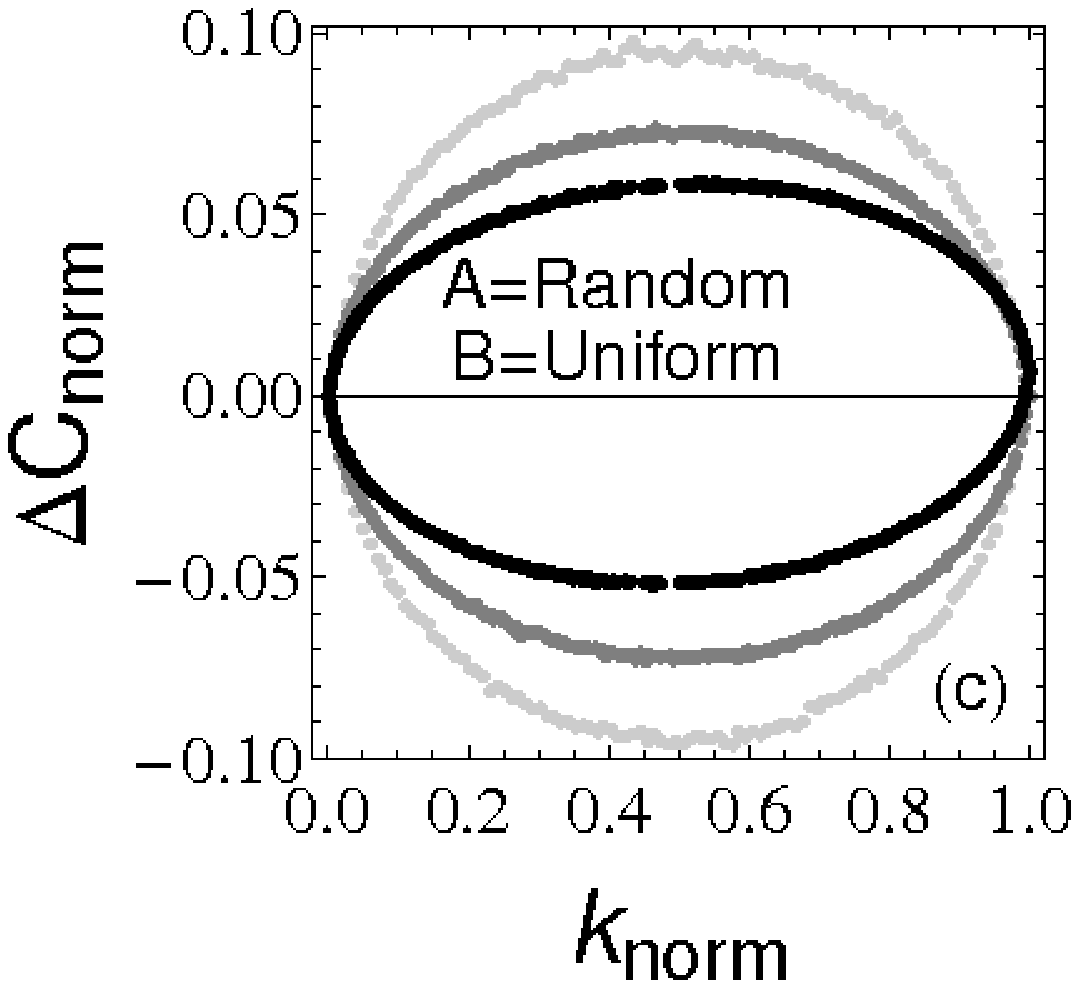} &
\epsfxsize=7.0cm
\epsfbox{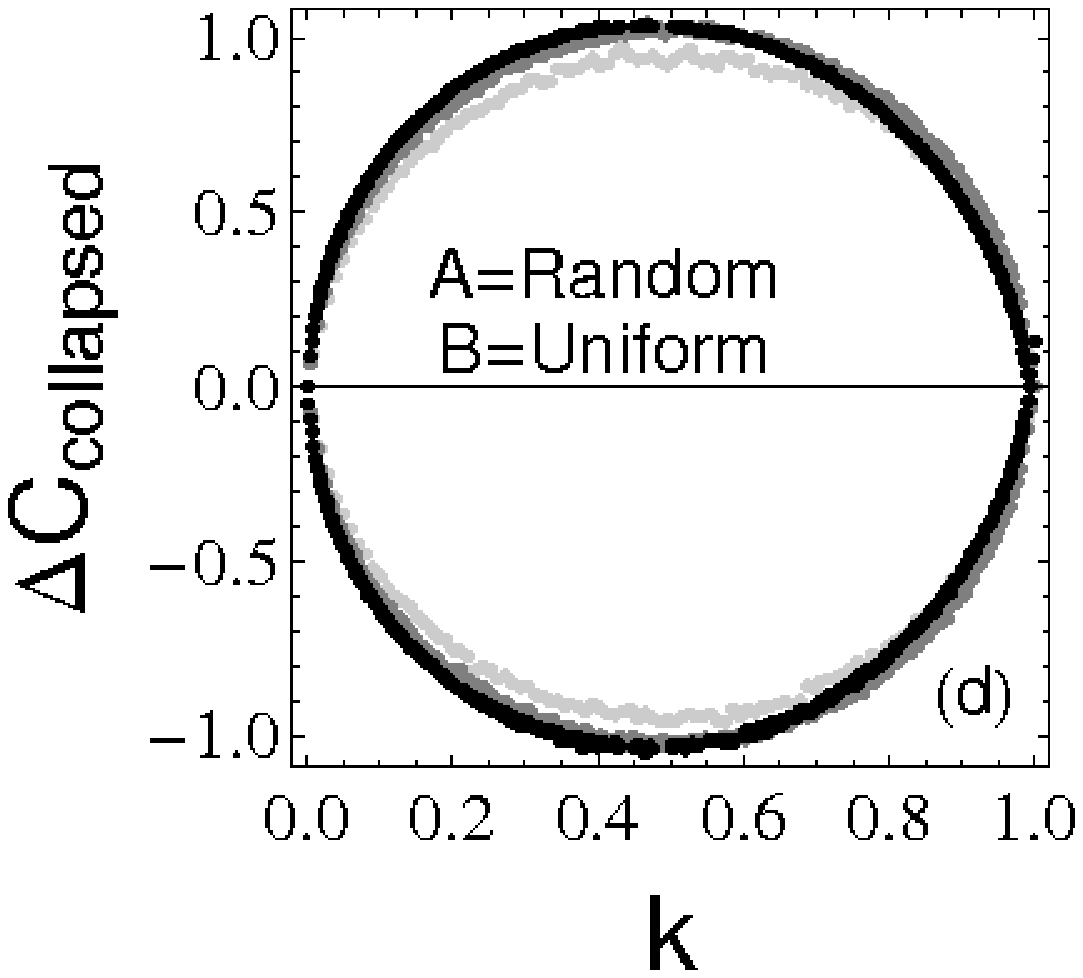} \\
\epsfxsize=7.0cm
\epsfbox{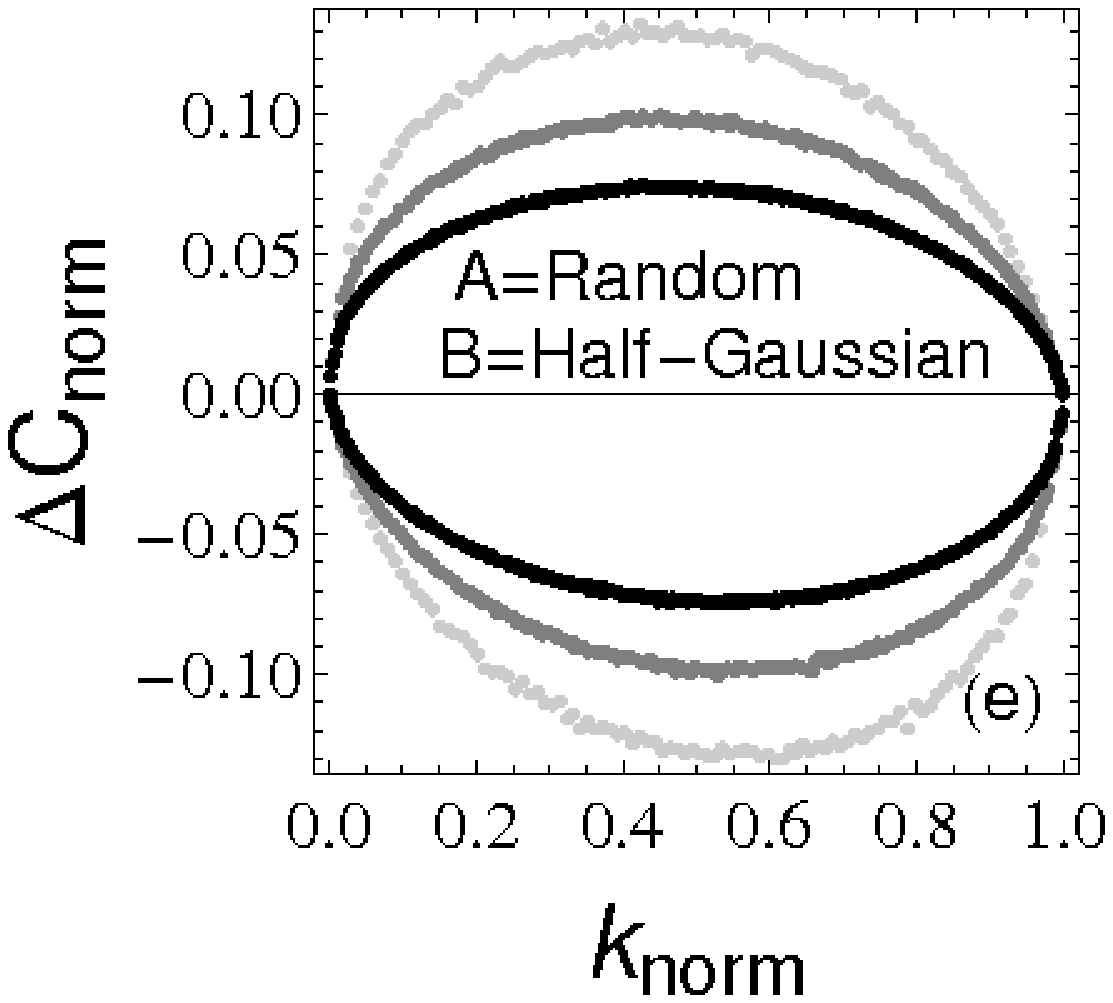} &
\epsfxsize=7.0cm
\epsfbox{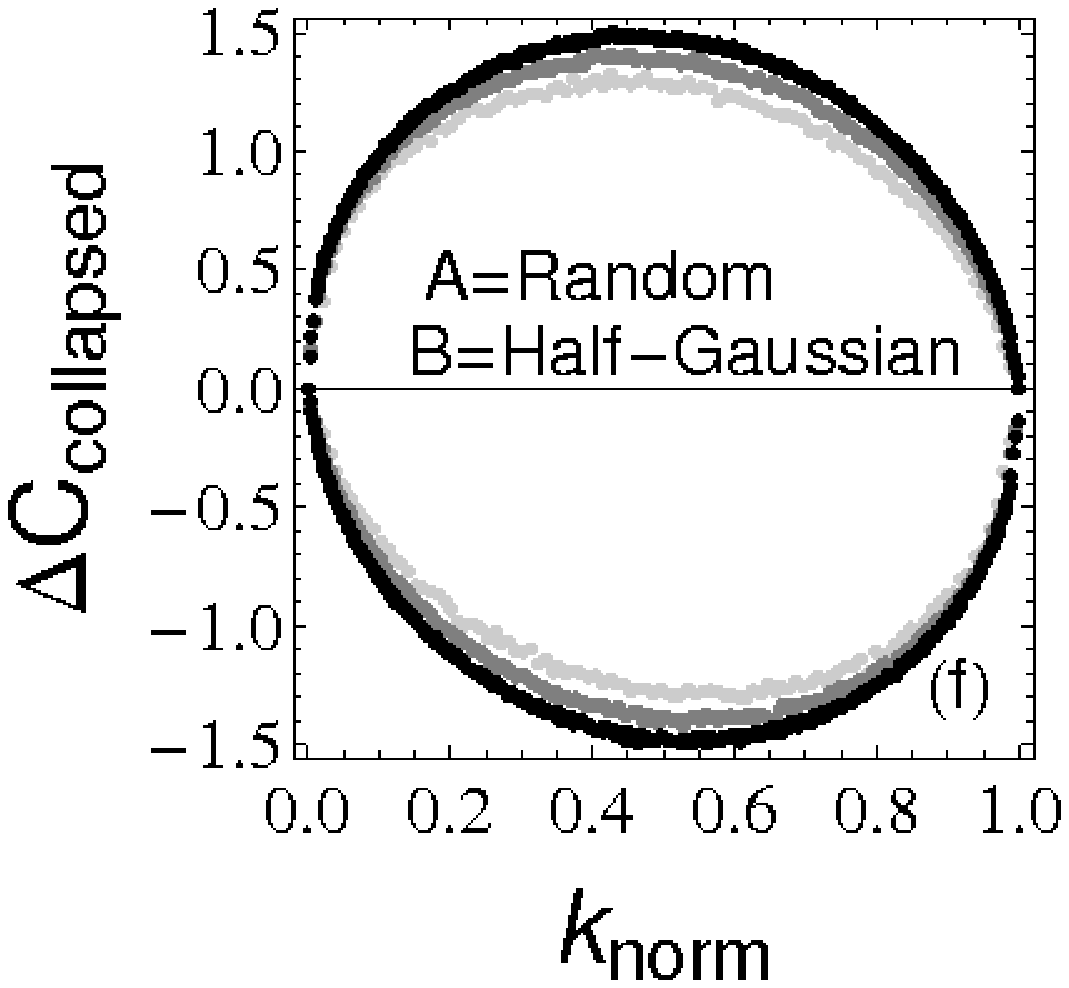} 
\end{array}$
\end{center}
\caption{(a),(c),(e) Normalized $\Delta C$ versus normalized $k$ for instance sizes $N$=100 (light gray); $ $N= 225 for (a )and  200 for (c) and (e) (medium gray); and $N$=400 (black).   The right hand column contains the  corresponding collapsed plots.
Panels (a) and (c) also appear in the main paper. }
\label{comb}
\end{figure}

\end{widetext}


\begin{thebibliography}{99}

\bibitem{dickey} Dickey, J. \& Hopkins, J. Campus building arrangement
  using TOPAZ. {\it Transportation Res.} {\bf 6}, 59--68 (1972).

\bibitem{carvalho} de Carvalho Jr., S. A. \& Rahmann, S. Microarray
  layout as a quadratic assignment problem. In Hudson, D. {\it et al.},
  eds. {\it German Conference on Bioinformatics (GCB)}, {\it Lecture
    Notes in Infomatics\/} {\bf P-83}, 11--20 (2006).

\bibitem{kolar} Kol\'a\v{r}, M., L\"assig, M. \& Berg, J.  From protein
  interactions to functional annotation: Graph alignment in Herpes. {\it
    BMC Systems Biol.} {\bf 2}, Article 90 (28 October 2008).

\bibitem{elshafei} Elshafei, A. N. Hospital layout as a quadratic
  assignment problem. {\it Operations Res. Quarterly\/} {\bf 28},
  167--179 (1977).

\bibitem{steinberg} Steinberg, L. The backboard wiring problem: A
  placement algorithm. {\it SIAM Rev.} {\bf 3}, 37--50 (1961).

\bibitem{Koopmans} Koopmans, T. \& Beckmann, M. Assignment problems and
  the location of economic activities. {\it Econometrica\/}, {\bf 25},
  53--76 (1957).

\bibitem{Anstreicher} Anstreicher, K.  Recent advances in the solution
  of quadratic assignment problems. {\it Math. Program\/} {\bf 97},
  27--42 (2003).

\bibitem{Cela} \c{C}ela, E. {\it The Quadratic Assignment Problem:
  Theory and Algorithms\/} (Kluwer, Boston, 1998).

\bibitem{James} James, T., Rego, C. \& Glover, F. Multistart tabu search
  and diversification strategies for the quadratic assignment
  problem. {\it IEEE Tran. on Systems, Man, and Cybernetics Part A:
    Systems and Humans\/} {\bf 39}, 579--596 (2009).


\bibitem{Loiola} E.M. Loiola, E. M. {\it et al.}
A survey for the quadratic assignment problem. {\it Eur. J. Operational
  Research\/} {\bf 176}, 657--690 (2007).

\bibitem{Pardalos} Pardalos, P. M., Rendl, F. \& Wolkowicz, H.  The
  quadratic assignment problem: A survey and recent developments.  In
  Pardalos, P. M. \& Wolkowicz, H., eds. {\it Quadratic Assignment and
    Related Problems: DIMACS Series on Discrete Mathematics and
    Theoretical Computer Science\/} {\bf 16} (Amer. Math. Soc.,
  Baltimore, MD, 1994), pp.~1--42.

\bibitem{hanan} Hanan, M. \& Kurtzberg, J. M., A review of the
  placement and quadratic assignment problems, SIAM Review {\bf 14}
  324-342 (1972).

\bibitem{liggett} Liggett R. S., The quadratic assignment problem: an
  analysis of applications and solution strategies, Environment and
  Planning B {\bf 7} (2) 141- 162 (1980).

\bibitem{nagarajan} Nagarajan, V., Sviridenko, M. On the Maximum
  Quadratic Assignment Problem, Mathematics of Operations Research
  {\bf 34}, No. 4, pp. 859-868 (2009).

\bibitem{makarychev} Makarychev, K., Manokaran, R., Sviridenko, M.
  Maximum Quadratic Assignment Problem: Reduction from Maximum Label
  Cover and LP-based Approximation Algorithm, In: Automata, Languages
  and Programming, 37th International Colloquium, Abramsky, S.,  Gavoille, C., Kirchner, C.  Meyer auf der Heide, F., Spirakis, P. G.
  (Eds.) ICALP 2010, Bordeaux, France, July 6-10, 2010, Proceedings,
  Part I. Lecture Notes in Computer Science 6198 Springer (2010)

\bibitem{burkard98} Burkard, R. E., Cela, E., Pardalos, P.M. \&
  Pitsoulis, L. S. The quadratic assignment problem, In Handbook of
  Combinatorial Optimization, Du, D.Z., Pardalos P. M. (Eds), {\bf 3}
  Kluwer Academic Publishers, 241-339 (1998).

\bibitem{burkard09} Burkard, R. E., Dell'Amico, M. \& Martello, S.
  Assignment Problems, SIAM Philadelphia (2009).

\bibitem{commander} Commander, C. W., A Survey of the Quadratic Assignment Problem, with Applications, Morehead Electronic Journal of Applicable 
Mathematics, {\bf 4} 1-15 (2005). 

\bibitem{burkardFincke} Burkard, R. E. \& Fincke, On random quadratic
  bottleneck assignment problems. {\it Math. Programming\/} {\bf 23},
  227--232 (1982).

\bibitem{frenk} Frenk, J. B. G., van Houweninge, M. \& Rinnooy, A. H. G.
  Asymptotic properties of the quadratic assignment problem. {\it
    Math. Oper. Res.} {\bf 10}, 100--116 (1985).

\bibitem{albrecher2005} Albrecher, H. A note on the asymptotic behaviour
  of bottleneck problems. {\it Oper. Res. Lett.} {\bf 33}, 183--186
  (2005).

\bibitem{albrecher2006} H. Albrecher, H., Burkard, R. E. \& \c{C}ela,
  E. J. An asymptotical study of combinatorial optimization problems by
  means of statistical mechanics. {\it J. Computational Appl. Math.}
  {\bf 186}, 148--162 (2006).

\bibitem{krokhmal} Krokhmal, P. \& Pardalos, P. Random assignment
  problems. {\it Eur. J. Operational Res.} {\bf 194}, 1--17 (2009).

\bibitem{EA} Edwards, S. F. \& Anderson, P. W. Theory of spin
  glasses. {\it J. Phys. F\/} {\bf 5}, 965--974 (1975).

\bibitem{SK} Kirkpatrick, S. \& Sherrington, D. Infinite-ranged models
  of spin-glasses. {\it Phys. Rev. B\/} {\bf 17}, 4384--4403 (1978).

\bibitem{sgtb} Mezard, M., Parisi, G. \& Virasoro, M. A. {\it Spin Glass
  Theory and Beyond\/} (World Scientific, Singapore, 1987).

\bibitem{Fu Anderson} Fu, Y. T. \& Anderson, P. W. Application of
  statistical-mechanics to NP-complete problems in combinatorial
  optimization. {\it J. Phys. A: Math. Gen.}  {\bf 19}, 1605--1620
  (1986).

\bibitem{hartmann} Hartmann, A. K. \& Weight, M. {\it Phase Transitions
  in Combinatorial Optimization Problems} (Wiley-VCH,
  Weinheim, 2005) Sec.~5.4.


\bibitem{Zdeborova} Zdeborov\'a, L. \& Boettcher, S. A conjecture on the
  maximum cut and bisection width in random regular graphs. {\it
    J. Stat. Mech.}  P02020 (2010).

\bibitem{Taillard} Taillard, \`E.  Robust taboo search for the quadratic
  assignment problem. {\it Parallel Comput.} {\bf 17}, 443--455 (1991).

\end{thebibliography}

\begin{thebibliography}{99}

\bibitem{EA} Edwards, S. F. \& Anderson, P. W. Theory of spin
  glasses. {\it J. Phys. F\/} {\bf 5}, 965--974 (1975).

\bibitem{SK} Kirkpatrick, S. \& Sherrington, D. Infinite-ranged models
  of spin-glasses. {\it Phys. Rev. B\/} {\bf 17}, 4384--4403 (1978).

\bibitem{sgtb} Mezard, M., Parisi, G. \& Virasoro, M. A. {\it Spin Glass
  Theory and Beyond\/} (World Scientific, Singapore, 1987).

\bibitem{Fu Anderson} Fu, Y. T. \& Anderson, P. W. Application of
  statistical-mechanics to NP-complete problems in combinatorial
  optimization. {\it J. Phys. A: Math. Gen.}  {\bf 19}, 1605--1620
  (1986).

\bibitem{hartmann} Hartmann, A. K. \& Weight, M. {\it Phase Transitions
  in Combinatorial Optimization Problems} (Wiley-VCH,
  Weinheim, 2005) Sec.~5.4.

\end{thebibliography}
\end{document}